\documentclass[preprint,12pt]{elsarticle}

\usepackage{amssymb}

\journal{***}

\begin{document}

\begin{frontmatter}



\title{Matrix-product structure of repeated-root constacyclic codes over finite fields}


\author{Yuan Cao$^{a}$, Yonglin Cao$^{a \ \ast}$, Fang-Wei Fu$^{b}$}

\address{$^{a}$School of Mathematics and Statistics,
Shandong University of Technology, Zibo, Shandong 255091, China
\vskip 1mm $^{b}$Chern Institute of Mathematics and LPMC, Nankai University, Tianjin 300071, China}
\cortext[cor1]{corresponding author.  \\
E-mail addresses:  yuancao@sdut.edu.cn (Yuan Cao), \ ylcao@sdut.edu.cn (Yonglin Cao),
 \ fwfu@nankai.edu.cn (F.--W. Fu).}

\begin{abstract}
For any prime number $p$, positive integers $m, k, n$
satisfying ${\rm gcd}(p,n)=1$ and $\lambda_0\in \mathbb{F}_{p^m}^\times$, we prove that any $\lambda_0^{p^k}$-constacyclic code of length $p^kn$ over the finite field $\mathbb{F}_{p^m}$ is monomially equivalent
to a matrix-product code of a nested sequence of $p^k$ $\lambda_0$-constacyclic codes with length $n$ over $\mathbb{F}_{p^m}$.
\end{abstract}


\end{frontmatter}


\section{Introduction}
\noindent
  Algebraic coding theory deals with the design of error-correcting and error-detecting codes for the reliable transmission
of information across noisy channel. The class of constacyclic codes play a very significant role in
the theory of error-correcting codes.

\par
  Let $R$ be a commutative finite ring with identity $1\neq 0$, and $R^{\times}$ be the multiplicative group of invertible elements of
$R$. For any $a\in
R$, we denote by $\langle a\rangle_R$, or $\langle a\rangle$ for
simplicity, the ideal of $R$ generated by $a$, i.e., $\langle
a\rangle_R=aR=\{ab\mid b\in R\}$. For any ideal $I$ of $R$, we will identify the
element $a+I$ of the residue class ring $R/I$ with $a$ (mod $I$) for
any $a\in R$.

\par
   A \textit{code}  of length $N$ over $R$ is a nonempty subset ${\cal C}$ of $R^N=\{(a_0,a_1,\ldots$, $a_{N-1})\mid a_j\in R, \
j=0,1,\ldots,N-1\}$. The code ${\cal C}$
is said to be \textit{linear} if ${\cal C}$ is an $R$-submodule of $R^N$. All codes in this paper are assumed to be linear.
   Let $\gamma\in R^{\times}$.
A linear code
${\cal C}$ of length $N$ over $R$ is
called a $\gamma$-\textit{constacyclic code}
if $(\gamma c_{N-1},c_0,c_1,\ldots,c_{N-2})\in {\cal C}$ for all
$(c_0,c_1,\ldots,c_{N-1})\in{\cal C}$. Particularly, ${\cal C}$ is
called a \textit{negacyclic code} if $\gamma=-1$, and ${\cal C}$ is
called a  \textit{cyclic code} if $\gamma=1$.

\par
  For any $a=(a_0,a_1,\ldots,a_{N-1})\in R^N$, let
$a(x)=a_0+a_1x+\ldots+a_{N-1}x^{N-1}\in R[x]/\langle x^N-\gamma\rangle$. We will identify $a$ with $a(x)$ in
this paper. It is well known that ${\cal C}$ is a  $\gamma$-constacyclic code
of length $N$ over $R$ if and only if ${\cal C}$ is an ideal of
the residue class ring $R[x]/\langle x^N-\gamma\rangle$. Now, let $R$ be a finite chain ring and $p$ be the characteristic of its residue class field. When ${\rm gcd}(p,N)=1$, ${\cal C}$ is called a
\textit{simple-root constacyclic code} while when $p\mid N$ it is called a \textit{repeated-root constacyclic code}.

\par
  For any positive integer $N$, we denote $[N)=\{0,1,\ldots,N-1\}$ in this paper. Let $C_1$ and $C_2$ be
codes of length $N$ over $R$. Recall that $C_1$ and $C_2$ are said to be \textit{monomially equivalent} if
there exists a permutation $\sigma$ on the set $[N)$ and fixed elements $r_0,r_1,\ldots,r_{N-1}\in R^\times$
such that
$$C_2=\{(r_0c_{\sigma(0)},r_1c_{\sigma(1)},\ldots,r_{N-1}c_{\sigma(N-1)})\mid (c_0,c_1,\ldots,c_{N-1})\in C_1\}$$
(cf. Huffman and  Pless [4] Page 24). Especially, $C_1$ and $C_2$ are said to be \textit{permutation equivalent} when $r_0=r_1=\ldots=r_{N-1}=1$ (cf. [4] Page 20).

\par
  From now on, let $\mathbb{F}_{p^m}$ be a finite field of cardinality $p^m$, where $p$ is a prime number, $m$, $k$  and $n$ are positive integers satisfying ${\rm gcd}(p,n)=1$ and $\lambda\in \mathbb{F}_{p^m}^\times$. In this paper, we concentrate on
$\lambda$-constacyclic codes of length $p^kn$ over $\mathbb{F}_{p^m}$, i.e.
ideals of the residue class ring $\mathbb{F}_{p^m}[x]/\langle x^{p^kn}-\lambda\rangle$. As ${\rm gcd}(p,n)=1$, there exits $\lambda_0\in \mathbb{F}_{p^m}^\times$ uniquely such that
$\lambda_0^{p^k}=\lambda.$

\par
    In this paper, we denote
$$\mathcal{R}:=\mathbb{F}_{p^m}[v]/\langle v^{p^k}-1\rangle=\mathbb{F}_{p^m}[v]/\langle (v-1)^{p^k}\rangle.$$
Set $u=v-1$. Then $\mathcal{R}=\mathbb{F}_{p^m}[u]/\langle u^{p^k}\rangle=\mathbb{F}_{p^m}+u\mathbb{F}_{p^m}+\ldots+u^{p^k-1}\mathbb{F}_{p^m}$ $(u^{p^k}=0)$, which is a finite
chain ring with the unique maximal ideal $u\mathcal{R}$, the nilpotency index of $u$ is equal to
$p^k$ and the residue class field $\mathcal{R}/u\mathcal{R}\cong \mathbb{F}_{p^m}$.

\par
  We will construct a precise isomorphism of rings from $\mathbb{F}_{p^m}[x]/\langle x^{p^kn}-\lambda\rangle$
onto $\mathcal{R}[x]/\langle x^n-\lambda_0\rangle$, which induces a one-to-one correspondence between the set
of  $\lambda$-constacyclic codes of length $p^kn$ over
$\mathbb{F}_{p^m}$ onto the set of $\lambda_0$-constascyclic codes of length $n$ over
$\mathcal{R}$.
   Paralleling to the theory of simple-root cyclic codes over $\mathcal{R}$, any $\lambda_0$-constascyclic code of length $n$ over
$\mathcal{R}$ can be determined uniquely by a tower of $\lambda_0$-constacyclic
codes with length $n$ over $\mathbb{F}_{p^m}$
$$\langle g_0(x)\rangle\subseteq \langle g_1(x)\rangle\subseteq\ldots\subseteq \langle g_{p^k-1}(x)\rangle\subseteq \mathbb{F}_{p^m}[x]/\langle x^n-\lambda_0\rangle,$$
where $g_0(x),g_1(x), \ldots, g_{p^k-1}(x)$ are monic divisors of $x^n-\lambda_0$ in $\mathbb{F}_{p^m}[x]$.
Then we give a direct description of the monomially equivalence between
a $\lambda$-constacyclic code of length $p^kn$ over $\mathbb{F}_{p^m}$ and a matrix-product code of the sequence of $\lambda_0$-constacyclic codes
$\langle g_s(x)\rangle$, $s=0,1,\ldots,p^k-1$.

\par
   In Section 2, we sketch the concept of matrix-product codes over finite fields and structures of simple-root constacyclic codes over the finite commutative chain ring $\mathcal{R}$. In Section 3, we prove that any $\lambda$-constacyclic code of length $p^kn$ over $\mathbb{F}_{p^m}$ is monomially equivalent
to a matrix-product code of a nested sequence of $p^k$ $\lambda_0$-constacyclic codes with length $n$ over $\mathbb{F}_{p^m}$. As an application, we determine the minimum Hamming distances
of all $4096$ negacyclic codes of length $7\cdot 2^l$ over $\mathbb{F}_7$
in Section 4 for any integer $l\geq 3$.



\section{Preliminaries}
\noindent
   We follow the notation in [2] Definition 2.1 for definition of matrix-product codes.
Let $A=[a_{ij}]$ be an $\alpha\times\beta$ matrix with entries in $\mathbb{F}_{p^m}$ and let
$C_1,\ldots,C_\alpha$ be codes of length $n$ over $\mathbb{F}_{p^m}$. The \textit{matrix-product code}
$[C_1,\ldots,C_\alpha]\cdot A$ is the set of all matrix products $[c_1,\ldots,c_\alpha]\cdot A$ defined by
\begin{eqnarray*}
[c_1,\ldots,c_\alpha]\cdot A&=&[c_1,\ldots,c_\alpha]\left[\begin{array}{cccc}a_{11} & a_{12} & \ldots & a_{1\beta}\cr
a_{21} & a_{22} & \ldots & a_{2\beta}\cr \vdots &\vdots &\vdots &\vdots \cr a_{\alpha 1} & a_{\alpha 2} & \ldots & a_{\alpha \beta}\end{array}\right]\\
 &=& [a_{11}c_1+a_{21}c_2+\ldots+a_{\alpha 1}c_{\alpha}, a_{12}c_1+a_{22}c_2+\ldots+a_{\alpha 2}c_{\alpha},\\
    && \ldots, a_{1\beta}c_1+a_{2\beta}c_2+\ldots+a_{\alpha \beta}c_{\alpha}]
\end{eqnarray*}
where $c_i\in C_i$ is an $n\times 1$ column vector for $1\leq i\leq \alpha$. Any codeword $[c_1,\ldots,c_\alpha]\cdot A$
is an $n\times \beta$ matrix over $\mathbb{F}_{p^m}$ and we regard it as a codeword of length $n\beta$ by reading the entries of
the matrix in column-major order.
A code $C$ over $\mathbb{F}_{p^m}$
is a matrix-product code if $C=[C_1,\ldots,C_\alpha]\cdot A$ for some codes
$C_1,\ldots,C_\alpha$ and a matrix $A$.

\par
  Given a matrix $A$ as above, let $A_t$ be the matrix consisting of the first $t$ rows of $A$.
For $1\leq j_1 < j_2 < \ldots < j_t \leq \beta$, we denote by $A(j_1, j_2, \ldots, j_t)$ the $t\times t$ submatrix consisting
of the columns $j_1, j_2, \ldots, j_t$ of $A_t$.  An $\alpha\times\beta$ matrix $A$ is said to be \textit{non-singular by columns} (NSC)
if every sub-matrix $A(j_1, j_2, \ldots, j_t)$ of $A$, $1\leq t\leq \alpha$, is non-singular (cf. [2] Definition 3.1).
Properties of matrix-product codes are described in the following result.

\vskip 3mm \noindent
  {\bf Theorem 2.1} (cf. [6], [3] Theorem 1)
\textit{The matrix-product code $[C_1, ..., C_\alpha] \cdot A$ given by a sequence
of $[n, k_i, d_i]$-linear codes $C_i$ over $\mathbb{F}_{p^m}$ and a full-rank matrix $A$ is a linear code whose length
is $n\beta$, it has dimension $\sum_{i=1}^\alpha k_i$
and minimum distance larger than or equal to
$$\delta={\rm min}\{d_i\delta_i\mid i=1,\ldots,\alpha\},$$
where $\delta_i$ is the minimum distance of the linear code with length $\beta$ over $\mathbb{F}_{p^m}$ generated by the first $i$ rows of
the matrix $A$. Moreover, when the matrix $A$ is NSC, it holds that
$\delta_i=\beta-i+1$. Furthermore, if we assume that the codes $C_i$ form a nested sequence
$C_1 \supseteq C_2 \supseteq \ldots\supseteq C_\alpha$, then the minimum distance of the code $[C_1 , C_2 , ..., C_\alpha] \cdot A$ is
exactly $\delta$.}

\vskip 3mm \noindent
  {\bf Theorem 2.2} (cf. [2] Proposition 6.2)
 \textit{Assume that $\{C_1, C_2, \ldots, C_s\}$ is a family of linear codes of length
$n$ and $A$ a non-singular $\alpha\times\alpha$ matrix, then the following equality of codes happens
$$\left([C_1, C_2, ..., C_\alpha] \cdot A\right)^{\bot}=[C_1^{\bot}, C_2^{\bot}, ..., C_\alpha^{\bot}] \cdot (A^{-1})^{\rm tr}$$
where $B^{\rm tr}$ denotes the transpose of the matrix $B$}.

\par
 \textit{Therefore, if $(A^{-1})^{\rm tr}=A$ and $C_i^{\bot}\subseteq C_i$ for all $i=1,\ldots,\alpha$, then}
$$\left([C_1, C_2, ..., C_\alpha] \cdot A\right)^{\bot}\subseteq [C_1, C_2, ..., C_\alpha] \cdot A.$$

\par
   Then we consider $\lambda_0$-constacyclic codes
of length $n$ over $\mathcal{R}=\mathbb{F}_{p^m}[u]/\langle u^{p^k}\rangle$ where $u=v-1$, i.e. ideals of the residue class ring
$\mathcal{R}[x]/\langle x^n-\lambda_0\rangle$. Let $\tau$ be a map defined
by
$\tau(b_0+b_1u+\ldots+b_{p^k-1}u^{p^k-1})=b_0$ for all $b_0, b_1, \ldots,b_{p^k-1}\in \mathbb{F}_{p^m}$.
Then $\tau$ is a surjective homomorphism of rings from $\mathcal{R}$ onto $\mathbb{F}_{p^m}$, which can be extended to
a surjective homomorphism of rings from $\mathcal{R}[x]/\langle x^n-\lambda_0\rangle$ onto $\mathbb{F}_{p^m}[x]/\langle x^n-\lambda_0\rangle$
in the natural way:
$$\alpha_0+\alpha_1 x+\ldots+\alpha_{n-1}x^{n-1}\mapsto \tau(\alpha_0)+\tau(\alpha_1) x+\ldots+\tau(\alpha_{n-1})x^{n-1},$$
for all $\alpha_0,\alpha_1,\ldots,\alpha_{n-1}\in \mathcal{R}$. We still use $\tau$ to denote this homomorphism.

\par
   Now, let $\mathcal{C}$ be a $\lambda_0$-constacyclic code
of length $n$ over $\mathcal{R}$. For any integer $s$, $0\leq s\leq p^k-1$, define
$$(\mathcal{C}:u^s)=\{\xi\in \mathcal{R}[x]/\langle x^n-\lambda_0\rangle\mid u^s\xi\in \mathcal{C}\}$$
which is an ideal of $\mathcal{R}[x]/\langle x^n-\lambda_0\rangle$ as well. It is clear that
\begin{equation}
\mathcal{C}=(\mathcal{C}:u^0)\subseteq (\mathcal{C}:u)\subseteq \ldots (\mathcal{C}:u^{p^k-1})
\subseteq (\mathcal{C}:u^{p^k})=\mathcal{R}[x]/\langle x^n-\lambda_0\rangle.
\end{equation}
Denote
$$T_s(\mathcal{C})=\tau((\mathcal{C}:u^s))=\{\tau(\xi)\mid \xi\in (\mathcal{C}:u^s)\}.$$
Then $T_s(\mathcal{C})$
is an ideal of the ring $\mathbb{F}_{p^m}[x]/\langle x^n-\lambda_0\rangle$. Hence there is a unique
monic divisor $g_s(x)$ of $x^n-\lambda_0$ in $\mathbb{F}_{p^m}[x]$ such that
$$T_s(\mathcal{C})=\langle g_s(x)\rangle=\{b(x)g_s(x)\mid {\rm deg}(b(x))<n-{\rm deg}(g_s(x)), \ b(x)\in \mathbb{F}_{p^m}[x]\},$$
where $g_s(x)$ is the generator polynomial of the $\lambda_0$-constacyclic
code $T_s(\mathcal{C})$. As $g_s(x)\in T_s(\mathcal{C})$, there exists $w_s(x)\in \mathcal{R}[x]$ such that $u^s(g_s(x)+u w_s(x))=u^sg_s(x)+u^{s+1}w_s(x)\in \mathcal{C}$. Then by $u^{p^k}=0$, we have
$$u^s(g_s(x))^{p^k-s}=u^s(g_s(x)+u w_s(x))\left(\sum_{t=0}^{p^k-s-1}g_s(x)^t\left(-uw_s(x)\right)^{p^k-s-1-t}\right)\in \mathcal{C}.$$
Since ${\rm gcd}(p,n)=1$, $x^n-\lambda_0$ has no repeated divisors in $\mathbb{F}_{p^m}[x]$, which implies
${\rm gcd}((g_s(x))^{p^k-s},x^n-\lambda_0)=g_s(x)$, and so there exist $a(x),b(x)\in \mathbb{F}_{p^m}[x]$ such
that $g_s(x)=a(x)(g_s(x))^{p^k-s}+b(x)(x^n-\lambda_0)$. Therefore, we have
\begin{equation}
u^sg_s(x)=a(x)\cdot u^s(g_s(x))^{p^k-s}\in \mathcal{C}, \ s=0,1,\ldots,p^k-1.
\end{equation}
By Equation (1), we have a tower of $\lambda_0$-constacyclic
codes
$$T_0(\mathcal{C})\subseteq T_1(\mathcal{C})\subseteq\ldots\subseteq T_{p^k-1}(\mathcal{C})\subseteq \mathbb{F}_{p^m}[x]/\langle x^n-\lambda_0\rangle,$$
which implies
$g_{p^k-1}(x)\mid\ldots \mid g_1(x)\mid g_0(x)\mid(x^n-\lambda_0)$ in $\mathbb{F}_{p^m}[x]$.

\par
   Let $c(x)\in \mathcal{C}$. Then $\tau(c(x))\in T_0(\mathcal{C})=\langle g_0(x)\rangle$. Hence there exists a unique polynomial $b_0(x)\in \mathbb{F}_{p^m}[x]$ satisfying ${\rm deg}(b_0(x))<n-{\rm deg}(g_0(x))$ such that $\tau(c(x))=b_0(x)g_0(x)$.
By Equation (2), it follows that $b_0(x)g_0(x)\in \mathcal{C}$, which implies
$c(x)-b_0(x)g_0(x)\in \mathcal{C}$.

\par
  As $\tau(c(x)-b_0(x)g_0(x))=\tau(c(x))-b_0(x)g_0(x)=0$, there exists $\alpha_1(x)\in \mathcal{R}[x]$ such that $u\alpha_1(x)=c(x)-b_0(x)g_0(x)\in \mathcal{C}$, which implies
$\alpha_1(x)\in (\mathcal{C}:u)$ and so $\tau(\alpha_1(x))\in T_1(\mathcal{C})$.
  By $T_1(\mathcal{C})=\langle g_1(x)\rangle$, there exists a unique polynomial $b_1(x)\in \mathbb{F}_{p^m}[x]$ satisfying ${\rm deg}(b_1(x))<n-{\rm deg}(g_1(x))$ such that $\tau(\alpha_1(x))=b_1(x)g_1(x)$. Then by Equation (2), it follows that
$ub_1(x)g_1(x)=b_1(x)\cdot ug_1(x)\in \mathcal{C}$. As $\tau(\alpha_1(x)-b_1(x)g_1(x))=0$, there exists $\alpha_2(x)\in \mathcal{R}[x]$ such that $u\alpha_2(x)=\alpha_1(x)-b_1(x)g_1(x)$ and
$$u^2\alpha_2(x)=u\alpha_1(x)-ub_1(x)g_1(x)\in \mathcal{C},$$
which implies
$\alpha_2(x)\in (\mathcal{C}:u^2)$ and so $\tau(\alpha_2(x))\in T_2(\mathcal{C})=\langle g_2(x)\rangle$.
  As stated above, we have
$$c(x)=b_0(x)g_0(x)+u\alpha_1(x)=b_0(x)g_0(x)+ub_1(x)g_1(x)+u^2\alpha_2(x),$$
where $b_0(x)g_0(x)\in T_0(\mathcal{C})$ and $b_1(x)g_1(x)\in T_1(\mathcal{C})$. Using mathematical induction on $s$, by Equation (2) and $u=v-1$ we have the following theorem.

\vskip 3mm \noindent
  {\bf Theorem 2.3}  \textit{Using the notations above, we have the following conclusions}.

\par
 (i) \textit{Let $\mathcal{C}$ be a $\lambda_0$-constacyclic code of length $n$ over $\mathcal{R}$. Then each
codeword $c(x)$ in $\mathcal{C}$ has a unique $(v-1)$-adic expansion}:
$$c(x)=\sum_{s=0}^{p^k-1}(v-1)^sc_s(x), \ {\rm where} \ c_s(x)\in T_s(\mathcal{C}), \ \forall s=0,1,\ldots,p^k-1.$$
\textit{Hence $|\mathcal{C}|=\prod_{s=0}^{p^k-1}|T_s(\mathcal{C})|=p^{m\sum_{s=0}^{p^k-1}(n-{\rm deg}(g_s(x)))}$}.

\par
 (ii) \textit{$\mathcal{C}$ is a $\lambda_0$-constacyclic code of length $n$ over $\mathcal{R}$ if and only if there exists uniquely
a tower of $\lambda_0$-constacyclic
codes of length $n$ over $\mathbb{F}_{p^m}$
$$C_0\subseteq C_1\subseteq\ldots\subseteq C_{p^k-1},$$
such that $T_s(\mathcal{C})=\tau(\mathcal{C}:(v-1)^s)=C_s$ for all $s=0,1,\ldots,p^k-1$}.

\vskip 3mm \noindent
 {\bf Remark} For a complete description of simple-root cyclic codes over arbitrary commutative finite chain rings,
please refer to [5] Theorem 3.5.



\section{The matrix-product structure of $\lambda$-constacyclic codes}
\noindent
First, we establish a precise relationship between the set of
$\lambda$-constacyclic codes of length $p^kn$ over the finite field $\mathbb{F}_{p^m}$ and
the set of $\lambda_0$-constacyclic codes of length $n$ over the finite chain ring $\mathcal{R}$.

\par
  Denote $[p^k)\times [n)=\{(s,j)\mid s\in [p^k), \ j\in [n)\}$. Then each $i\in [p^kn)$ can be uniquely expressed as
\begin{equation}
i=sn+j, \ {\rm were} \ s=\left\lfloor \frac{i}{n}\right\rfloor\in [p^k) \ {\rm and} \ j=i-sn\in [n).
\end{equation}

\par
  For any $a(x)=\sum_{j=0}^{n-1}\sum_{s=0}^{p^k-1}a_{j+sn}x^{j+sn}\in \mathbb{F}_{p^m}[x]/\langle x^{p^kn}-\lambda\rangle$, by Equation (3)
$a(x)$ can be uniquely expressed as the following product of matrices
\begin{equation}
a(x)=[1,x,x^{2},\ldots,x^{n-1}]M_{a(x)}\Xi
\end{equation}
where $\Xi=[1,x^n,x^{2n},\ldots,x^{(p^k-1)n}]^{{\rm tr}}$ is the transpose of the $1\times p^k$ matrix
$[1,x^n,x^{2n},\ldots,x^{(p^k-1)n}]$ and
$$M_{a(x)}=\left[\begin{array}{ccccc}a_{0} & a_{n} & a_{2n} &\ldots & a_{(p^k-1)n} \cr
 a_{1} & a_{1+n} & a_{1+2n} &\ldots & a_{1+(p^k-1)n} \cr \ldots  &\ldots  & \ldots  & \ldots & \ldots  \cr
 a_{n-1} & a_{n-1+n} & a_{n-1+2n} &\ldots & a_{n-1+(p^k-1)n}
\end{array}\right]$$
is a $n\times p^k$ matrix over $\mathbb{F}_{p^m}$.
  As $\lambda_0^{p^k}=\lambda$, we have $(\frac{x^n}{\lambda_0})^{p^k}=\frac{x^{p^kn}}{\lambda}=1$ in $\mathbb{F}_{p^m}[x]/\langle x^{p^kn}-\lambda\rangle$. Now, set
\begin{equation}
v=\frac{x^n}{\lambda_0}, \ {\rm i.e.} \ x^n=\lambda_0v,  \ {\rm and} \
D_{\lambda_0}={\rm diag}(1,\lambda_0,\lambda_0^2,\ldots,\lambda_0^{p^k-1}).
\end{equation}
Then by Equation (4) it follows that
\begin{eqnarray*}
a(x)&=&[1,x,x^{2},\ldots,x^{n-1}]M_{a(x)}D_{\lambda_0}
 \left[1,\frac{x^n}{\lambda_0},(\frac{x^n}{\lambda_0})^2,\ldots,(\frac{x^n}{\lambda_0})^{(p^k-1)}\right]^{{\rm tr}}\\
  &=&[1,x,x^{2},\ldots,x^{n-1}]M_{a(x)}D_{\lambda_0}V.
\end{eqnarray*}
where $V=[1,v,v^2,\ldots,v^{p^k-1}]^{{\rm tr}}$ is the transpose of the $1\times p^k$ matrix
$[1,v,v^2$, $\ldots,v^{p^k-1}]$.
  We define a map
$\phi: \mathbb{F}_{p^m}[x]/\langle x^{p^kn}-\lambda\rangle\rightarrow \mathcal{R}/\langle x^n-\lambda_0v\rangle$
by
\begin{eqnarray*}
\phi(a(x))&=&[1,x,x^{2},\ldots,x^{n-1}]\left(M_{a(x)}D_{\lambda_0}V\right)\\
 &=&\alpha_0+\alpha_1 x+\ldots+\alpha_{n-1} x^{n-1}
\end{eqnarray*}
where $[\alpha_0,\alpha_1,\ldots,\alpha_{n-1}]^{{\rm tr}}=M_{a(x)}D_{\lambda_0}V\in \mathcal{R}^{n}$. Then one
can easily verify the following conclusion. We omit the proof here.

\vskip 3mm \noindent
  {\bf Lemma 3.1} \textit{The map $\phi$ defined above is an isomorphism of rings from $\mathbb{F}_{p^m}[x]/\langle x^{p^kn}-\lambda\rangle$
onto $\mathcal{R}/\langle x^n-\lambda_0v\rangle$}.

\vskip 3mm \par
  As ${\rm gcd}(p,n)=1$ and $v^{p^k}=(\frac{x^n}{\lambda_0})^{p^k}=1$ in $\mathcal{R}/\langle x^n-\lambda_0v\rangle$, there exists a unique integer $n^\prime$,
$0\leq n^\prime \leq p^k-1$, such that $n^\prime n\equiv 1$ (mod $p^k$), which implies
$(v^{n^\prime})^n=v^{n^\prime n}=v$. Define a map $\psi: \mathcal{R}/\langle x^n-\lambda_0v\rangle\rightarrow \mathcal{R}/\langle x^n-\lambda_0\rangle$ by
$$\alpha(x)\mapsto\alpha(v^{n^\prime}x)=[1,x,\ldots,x^{n-1}]{\rm diag}(1,v^{n^\prime},\ldots, (v^{n^\prime})^{n-1})[\alpha_0,\alpha_1,\ldots,\alpha_{n-1}]^{{\rm tr}}$$
for any $\alpha(x)=\alpha_0+\alpha_1 x+\ldots+\alpha_{n-1} x^{n-1}\in \mathcal{R}/\langle x^n-\lambda_0v\rangle$.
As $\psi(x^n)=(v^{n^\prime}x)^n=vx^n$ and $v\in\mathcal{R}^\times$, it can be verify easily that $\psi$ is an isomorphism of rings
from $\mathcal{R}/\langle x^n-\lambda_0v\rangle$ onto $\mathcal{R}/\langle x^n-\lambda_0\rangle$.

\par
  Then by Lemma 3.1, we conclude the following conclusion.

\vskip 3mm \noindent
  {\bf Lemma 3.2} \textit{Using the notations above, the map $\psi\phi$ define by
$$\psi\phi(a(x))=[1,x,\ldots,x^{n-1}]{\rm diag}(1,v^{n^\prime},\ldots, (v^{n^\prime})^{n-1})M_{a(x)}D_{\lambda_0}V$$
$(\forall a(x)\in \mathbb{F}_{p^m}[x]/\langle x^{p^kn}-\lambda\rangle)$ is an isomorphism
of rings from $\mathbb{F}_{p^m}[x]/\langle x^{p^kn}-\lambda\rangle$ onto $\mathcal{R}/\langle x^n-\lambda_0\rangle$. Therefore,
$C$ is a $\lambda$-constacyclic code of length $p^kn$ over $\mathbb{F}_{p^m}$ if and only if
$\psi(\phi(C))$ is a $\lambda_0$-constacyclic code of length $n$ over $\mathcal{R}$}.

\vskip 3mm \par
  Now, using Lemma 3.2 and Theorem 2.3 we give the matrix-product structure of any $\lambda$-constacyclic code of length $p^kn$ over $\mathbb{F}_{p^m}$.

\vskip 3mm \noindent
  {\bf Theorem 3.3} \textit{Using the notations above, let $C$ be a $\lambda$-constacyclic code of length $p^kn$ over $\mathbb{F}_{p^m}$, assume $\mathcal{C}=\psi(\phi(C))$ and $C_s=T_s(\mathcal{C})$ for all $s=0,1,\ldots,p^k-1$. Then
$C$ is monomially equivalent to the matrix-product code
$$[C_{p^k-1},\ldots, C_1,C_0]\cdot A$$
of the nested sequences $C_{p^k-1}\supseteq\ldots\supseteq C_1\supseteq C_0$ which are $\lambda_0$-constacyclic codes of length $n$ over $\mathbb{F}_{p^m}$, where}
{\scriptsize
$$A=\left[\begin{array}{cccccc} (-1)^{p^k-1} & (-1)^{p^k-2}\left(\begin{array}{c}p^k-1 \cr 1\end{array}\right)
& (-1)^{p^k-3}\left(\begin{array}{c}p^k-1 \cr 2\end{array}\right) & \ldots & (-1)\left(\begin{array}{c}p^k-1 \cr p^k-2\end{array}\right) & 1 \cr
(-1)^{p^k-2} & (-1)^{p^k-3}\left(\begin{array}{c}p^k-2 \cr 1\end{array}\right)
&  (-1)^{p^k-4}\left(\begin{array}{c}p^k-2 \cr 2\end{array}\right) & \ldots  & 1 & 0\cr
\ldots & \ldots &  \ldots & \ldots  & 0& 0 \cr
(-1)^2 & (-1)\left(\begin{array}{c}2 \cr 1\end{array}\right) & 1 & \ldots & 0 & 0\cr
-1 & 1 & 0 &\ldots & 0 & 0\cr
1 & 0 & 0 & \ldots & 0 & 0
\end{array}\right].$$}

\vskip 3mm \noindent
  {\bf Proof.} Let $a(x)=\sum_{j=0}^{n-1}\sum_{s=0}^{p^k-1}a_{j,s}x^{sn+j}\in C$ where $a_{i,j}\in \mathbb{F}_{p^m}$, and assume $\beta(x)=\psi\phi(a(x))\in \mathcal{C}$. For each integer $s$, $0\leq s\leq p^k-1$, by Theorem 2.3 there exists a unique codeword $c_s(x)\in C_s$ such that
\begin{eqnarray*}
\beta(x)&=&c_0(x)+(v-1)c_1(x)+(v-1)^2c_2(x)+\ldots+(v-1)^{p^k-1}c_{p^k-1}(x)\\
        &=&[c_{p^k-1}(x),\ldots,c_1(x),c_0(x)]\left[\begin{array}{c}(v-1)^{p^k-1}\cr \ldots\cr v-1\cr 1\end{array}\right]\\
        &=&[c_{p^k-1}(x),\ldots,c_1(x),c_0(x)]AV\\
        &=&[1,x,\ldots,x^{n-1}][\textbf{c}_{p^k-1},\ldots,\textbf{c}_1,\textbf{c}_0]AV
\end{eqnarray*}
where $V=[1,v,v^2,\ldots,v^{p^k-1}]^{{\rm tr}}$ is the transpose of $[1,v,v^2,\ldots,v^{p^k-1}]$ and
$c_s(x)=[1,x,\ldots,x^{n-1}]\textbf{c}_s$ in which $\textbf{c}_s\in C_s$ being a $n\times 1$ column vector over $\mathbb{F}_{p^m}$ for
all $s=0,1,\ldots, p^k-1$. Replacing $v$ with $\frac{x^n}{\lambda_0}$, we obtain
$$
\pi(\beta(x))=[1,x,\ldots,x^{n-1}][\textbf{c}_{p^k-1},\ldots,\textbf{c}_1,\textbf{c}_0]A
\left[1,\frac{x^n}{\lambda_0},\ldots,(\frac{x^n}{\lambda_0})^{p^k-1}\right]^{{\rm tr}}.$$
By $D_{\lambda_0}={\rm diag}(1,\lambda_0,\ldots,\lambda_0^{p^k-1})$ and $\Xi=[1,x^n,\ldots,x^{(p^k-1)n}]^{{\rm tr}}$, we have
\begin{equation}
\pi(\beta(x))=[1,x,\ldots,x^{n-1}]\left([\textbf{c}_{p^k-1},\ldots,\textbf{c}_1,\textbf{c}_0]AD_{\lambda_0}^{-1}\right)\Xi
\end{equation}

\par
   On the other hand, by Lemma 3.2 we have
$$\beta(x)=[1,x,\ldots,x^{n-1}]{\rm diag}(1,v^{n^\prime},\ldots, (v^{n^\prime})^{n-1})M_{a(x)}D_{\lambda_0}V.$$
Replacing $v$ with $\frac{x^n}{\lambda_0}$, by $V=D_{\lambda_0}^{-1}\Xi$ we obtain
\begin{eqnarray*}
\pi(\beta(x))&=&[1,x,\ldots,x^{n-1}]{\rm diag}(1,(\frac{x^n}{\lambda_0})^{n^\prime},\ldots, ((\frac{x^n}{\lambda_0})^{n^\prime})^{n-1})M_{a(x)}D_{\lambda_0}D_{\lambda_0}^{-1}\Xi\\
 &=&[1,x^{1+n^\prime n},x^{2(1+n^\prime n)},\ldots,x^{(n-1)(1+n^\prime n)}]\\
  &&\cdot {\rm diag}(1,(\frac{1}{\lambda_0})^{n^\prime},(\frac{1}{\lambda_0})^{2n^\prime},\ldots, (\frac{1}{\lambda_0})^{(n-1)n^\prime})M_{a(x)}\Xi\\
  &=&\sum_{j=0}^{n-1}\sum_{s=0}^{p^k-1}(\frac{1}{\lambda_0^{n^\prime}})^ja_{j+sn}x^{j(1+n^\prime n)+sn},
\end{eqnarray*}
where $j(1+n^\prime n)+sn=(s+n^\prime j)n+j$. Denote $t=s+j n^\prime $ (mod $p^k$). Then
$s=t-j n^\prime$ (mod $p^k$) and hence
\begin{eqnarray*}
\pi(\beta(x))&=&\sum_{j=0}^{n-1}\sum_{t=0}^{p^k-1}
(\frac{1}{\lambda_0^{n^\prime}})^ja_{j+(t-j n^\prime)n}x^{j+tn}\\
 &=&[1,x,\ldots,x^{n-1}]\left({\rm diag}(1,(\frac{1}{\lambda_0})^{n^\prime},\ldots, ((\frac{1}{\lambda_0})^{n^\prime})^{n-1})\widehat{M}_{a(x)}\right)\Xi,
\end{eqnarray*}
where $\widehat{M}_{a(x)}=\left[\widehat{a}_{j,t}\right]_{0\leq j\leq n-1, 0\leq t\leq p^k-1}$ is a $n\times p^k$ matrix over $\mathbb{F}_{p^k}$ with entries
\begin{equation}
\widehat{a}_{j,t}=a_{j+(t-j n^\prime)n},
\ 0\leq j\leq n-1, \ 0\leq t\leq p^k-1.
\end{equation}

\par
   As stated above, by Equations (6) and (7) it follows that
$${\rm diag}(1,(\frac{1}{\lambda_0})^{n^\prime},\ldots, ((\frac{1}{\lambda_0})^{n^\prime})^{n-1})\widehat{M}_{a(x)}=[\textbf{c}_{p^k-1}, \ldots,\textbf{c}_1,\textbf{c}_0]AD_{\lambda_0}^{-1}.$$
This implies
\begin{equation}
{\rm diag}(1,(\frac{1}{\lambda_0})^{n^\prime},\ldots, ((\frac{1}{\lambda_0})^{n^\prime})^{n-1})\widehat{M}_{a(x)}D_{\lambda_0}
=[\textbf{c}_{p^k-1},\ldots,\textbf{c}_1,\textbf{c}_0]A.
\end{equation}
As $(t-n^\prime j,j)=(t,j)\left[\begin{array}{cc} 1& 0\cr -n^\prime & 1\end{array}\right]$
where $t-n^\prime j$ modulo $p^k$, the following formula
$$\sigma(j+tn)=j+n(t-n^\prime j \ \left({\rm mod} \ p^k)\right) \ {\rm for} \ {\rm any} \ j\in [n)
\ {\rm and} \ t\in [p^k), $$
defines a permutation $\sigma$ on the set $[p^kn)=\{0,1,\ldots,p^kn-1\}$. From this and by Equation (8),
we deduce that
$$\left[\begin{array}{ccccc}b_{0} & b_{n} & b_{2n} &\ldots & b_{(p^k-1)n} \cr
 b_{1} & b_{1+n} & b_{1+2n} &\ldots & b_{1+(p^k-1)n} \cr \ldots  &\ldots  & \ldots  & \ldots & \ldots  \cr
 b_{n-1} & b_{n-1+n} & b_{n-1+2n} &\ldots & b_{n-1+(p^k-1)n}
\end{array}\right]=[\textbf{c}_{p^k-1},\ldots,\textbf{c}_1,\textbf{c}_0]A$$
in which
$$b_{j+tn}=(\frac{1}{\lambda_0^{n^\prime}})^j\widehat{a}_{j,t}\lambda_0^t
=\lambda_0^{t-jn^\prime}a_{\sigma(j+tn)} \ {\rm for} \ {\rm all} \ j\in [n)
\ {\rm and} \ t\in [p^k).$$
This implies that
$C$ is monomially equivalent to the matrix-product code $[C_{p^k-1},\ldots$, $C_1,C_0]\cdot A$ of
$\lambda_0$-constacyclic codes $C_{p^k-1}\supseteq\ldots\supseteq C_1\supseteq C_0$.
\hfill
$\Box$

\vskip 3mm\noindent
  {\bf Remark} When $\lambda=1$, we have $\lambda_0=1$ and that Equation (8) is simplified to
$$\widehat{M}_{a(x)}=[\textbf{c}_{p^k-1},\ldots,\textbf{c}_1,\textbf{c}_0]A,$$
where $\textbf{c}_s\in C_s$ and $C_s$ is a cyclic code of length $n$ over $\mathbb{F}_{p^m}$
for all $s=0,1,\ldots, p^k-1$.
Hence any cyclic code of length $p^kn$ over $\mathbb{F}_{p^m}$ is permutation equivalent
to a matrix-product code of a nested sequence of $p^k$ cyclic codes with length $n$ over $\mathbb{F}_{p^m}$. This conclusion was proved by Sobhani [7].

\vskip 3mm \par
   In order to get properties of a $\lambda$-constacyclic code $C$
of length $p^kn$ using Theorem 3.3, we need to determine the nested sequences $C_{p^k-1}\supseteq\ldots\supseteq C_1\supseteq C_0$ of
$\lambda_0$-constacyclic codes with length $n$.   As ${\rm gcd}(p,n)=1$, we have that $x^n-\lambda_0=f_1(x)f_2(x)\ldots f_r(x)$ where $f_1(x),f_2(x),\ldots, f_r(x)$
are pairwise coprime monic irreducible polynomials in $\mathbb{F}_{p^m}[x]$, which implies
$$x^{p^kn}-\lambda=(x^n-\lambda_0)^{p^k}=f_1(x)^{p^k}f_2(x)^{p^k}\ldots f_r(x)^{p^k}.$$
Recall that each $\lambda$-constacyclic code of length $p^kn$ over $\mathbb{F}_{p^m}$ has a unique
monic divisor $g(x)$ of $x^{p^kn}-\lambda$ in $\mathbb{F}_{p^m}[x]$ as its generator polynomial.

\vskip 3mm \noindent
  {\bf Theorem 3.4} \textit{Let $C$ be a  $\lambda$-constacyclic code $C$ of length $p^kn$ over $\mathbb{F}_{p^m}$
with generator polynomial $g(x)=f_1(x)^{i_1}f_2(x)^{i_2}\ldots f_r(x)^{i_r}$, where $0\leq i_1,i_2,\ldots,i_r$ $\leq p^k$.
Then $C$ is monomially equivalent to the matrix-product code $[C_{p^k-1}$, $\ldots, C_1,C_0]\cdot A$, where $A$ is given by Theorem 3.3
and for each integer $s$, $0\leq s\leq p^k-1$, $C_s$ is a $\lambda_0$-constacyclic code of length $n$
over $\mathbb{F}_{p^m}$ with generator polynomial}
 $$g_s(x)=\prod_{i_t>s, \ 1\leq t\leq r}f_t(x).$$

\vskip 3mm \noindent
  {\bf Proof.} Let $0\leq s\leq p^k-1$. It is suffices to prove that $C_s=\langle g_s(x)\rangle$.
To do this, we first verify that $g_s(x)\in C_s=T_s(\psi(\phi(C)))=\tau(\psi(\phi(C)):(v-1)^s)$, which is equivalent to that
\begin{equation}
(v-1)^s\left(g_s(x)+(v-1)q(x)\right)\in \psi(\phi(C))
\end{equation}
for some $q(x)\in \mathcal{R}[x]/\langle x^n-\lambda_0\rangle$. Now, we denote
$$A_s=\{t\mid i_t>s, \ 1\leq t\leq r\}, \  B_s=\{t\mid i_t\leq s, \ 1\leq t\leq r\}.$$
and set
$$h_s(x)=\prod_{t\in B_s}f_t(x), \ \widehat{f}_s(x)=\prod_{t\in A_s}f_t(x)^{i_t-s-1}.$$
Then $g_s(x)=\prod_{t\in A_s}f_t(x)$, $x^n-\lambda_0=g_s(x)h_s(x)$,
${\rm gcd}(g_s(x),h_s(x))=1$ and  ${\rm gcd}(\widehat{f}_s(x),h_s(x))=1$. By
$g(x)=\prod_{t\in A_s\cup B_s}f_t(x)^{i_t}$ and $i_t\leq s$ for all $t\in B_s$, we have
$$(x^n-\lambda_0)^sg_s(x)\widehat{f}_s(x)=\prod_{t\in A_s\cup B_s}f_t(x)^s\prod_{t\in A_s}f_t(x)\prod_{t\in A_s}f_t(x)^{i_t-s-1}
=e(x)g(x)$$
where $e(x)=\prod_{t\in B_s}f_t(x)^{s-i_t}\in\mathbb{F}_{p^m}[x]$, which implies
\begin{equation}
(x^n-\lambda_0)^sg_s(x)\widehat{f}_s(x)\in \langle g(x)\rangle=C.
\end{equation}
As ${\rm gcd}(\widehat{f}_s(x),h_s(x))=1$, there exist $a(x),b(x)\in \mathbb{F}_{p^m}[x]$ such that
$a(x)\widehat{f}_s(x)+b(x)h_s(x)=1$, i.e., $a(x)\widehat{f}_s(x)=1-b(x)h_s(x)$. Then by Equation (10) it follows
that
\begin{eqnarray*}
&&(x^n-\lambda_0)^sg_s(x)-(x^n-\lambda_0)^{s+1}b(x)\\
 &=&(x^n-\lambda_0)^sg_s(x)-(x^n-\lambda_0)^{s}\cdot g_s(x)h_s(x)\cdot b(x)\\
 &=&(x^n-\lambda_0)^sg_s(x)(1-b(x)h_s(x))\\
 &=&(x^n-\lambda_0)^sg_s(x)\widehat{f}_s(x)\cdot a(x)\in C.
\end{eqnarray*}
We replace $x^n$ with $\lambda_0v$ as in (5), by the definition of $\phi$, we obtain
\begin{eqnarray*}
&&\lambda_0^s\left((v-1)^sg_s(x)-(v-1)^{s+1}\lambda_0\phi(b(x))\right)\\
 &=&(\lambda_0v-\lambda_0)^sg_s(x)-(\lambda_0v-\lambda_0)^{s+1}\phi(b(x))\\
 &=&\phi\left((x^n-\lambda_0)^sg_s(x)-(x^n-\lambda_0)^{s+1}b(x)\right)\in \phi(C),
\end{eqnarray*}
which implies $(v-1)^sg_s(x)-(v-1)^{s+1}\alpha(x)\in \phi(C)$, where $\alpha(x)=\lambda_0\phi(b(x))\in
\mathcal{R}[x]/\langle x^n-\lambda_0v\rangle$. Then we replace $x$ with $v^{n^\prime}x$, by the definition
of $\psi$ we have
\begin{eqnarray*}
&&(v-1)^sg_s(v^{n^\prime}x)-(v-1)^{s+1}\alpha(v^{n^\prime}x)\\
&=&\psi\left(v-1)^sg_s(x)-(v-1)^{s+1}\alpha(x)\right)\in \psi(\phi(C)).
\end{eqnarray*}
From this and by
$$g_s(v^{n^\prime}x)=g_s(x+x(v^{n^\prime}-1))=g_s(x+(v-1)\beta(x))=g_s(x)+(v-1)\gamma(x)$$
for some $\gamma(x)\in \mathcal{R}[x]/\langle x^n-\lambda_0\rangle$, where $\beta(x)=x\sum_{i=0}^{n^\prime-1}v^i$, we deduce that
\begin{eqnarray*}
&&(v-1)^s\left(g_s(x)+(v-1)(\gamma(x)-\alpha(v^{n^\prime}x))\right)\\
&=&(v-1)^s(g_s(x)+(v-1)\gamma(x))-(v-1)^{s+1}\alpha(v^{n^\prime}x)\in \psi(\phi(C)),
\end{eqnarray*}
which implies $g_s(x)+(v-1)(\gamma(x)-\alpha(v^{n^\prime}x))\in (\psi(\phi(C)):(v-1)^s)$, and so
$g_s(x)=\tau\left(g_s(x)+(v-1)(\gamma(x)-\alpha(v^{n^\prime}x))\right)\in \tau(\psi(\phi(C)):(v-1)^s)=C_s$.

\par
  As stated above, we conclude that $\langle g_s(x)\rangle\subseteq C_s$ as ideals in $\mathbb{F}_{p^m}[x]/\langle x^n-\lambda_0\rangle$
for all $s=0,1,\ldots,p^k-1$. By Theorem 2.3 and the properties of constacyclic codes over finite fields,
we have
\begin{eqnarray*}
(p^m)^{p^kn-{\rm deg}(g(x))}&=&|C|=|\psi(\phi(C))|=\prod_{s=0}^{p^k-1}|C_s|\\
  &\geq&\prod_{s=0}^{p^k-1}|\langle g_s(x)\rangle|=(p^m)^{\sum_{s=0}^{p^k-1}(n-{\rm deg}(g_s(x)))}.
\end{eqnarray*}
From this and by $g(x)=\prod_{s=0}^{p^k-1}g_s(x)$, we deduce that $|C_s|=|\langle g_s(x)\rangle|$. Hence
$C_s=\langle g_s(x)\rangle$, i.e. $g_s(x)$ is the generator polynomial of $C_s$ for all $s$.
\hfill
$\Box$


\section{An Example} \label{}
  We consider
negacyclic codes of length $7\cdot 2^l$ over $\mathbb{F}_7$ where
$l$ is an integer satisfying $l\geq 3$.
  It is known that $x^{2^l}+1=
f_1(x)f_2(x)f_3(x)f_4(x)$ where

\par
$f_1(x)=x^{2^{l-2}}+x^{2^{l-3}}+6$, $f_2(x)=x^{2^{l-2}}+6x^{2^{l-3}}+6$,

\par
$f_3(x)=x^{2^{l-2}}+4x^{2^{l-3}}+6$, $f_4(x)=x^{2^{l-2}}+3x^{2^{l-3}}+6$,

\noindent
which are irreducible polynomials in $\mathbb{F}_7[x]$ (see [1], for example).

\par
  As $(x^{2^l}+1)^7=f_1(x)^7f_2(x)^7f_3(x)^7f_4(x)^7$, all distinct $4096$ negacyclic
codes over $\mathbb{F}_7$ of length $7\cdot2^l$ are given by
$$\mathcal{C}_{(j_1,j_2,j_3,j_4)}=\langle f_1(x)^{j_1}f_2(x)^{j_2}f_3(x)^{j_3}f_4(x)^{j_4}\rangle,
\ 0\leq j_1,j_2,j_3,j_4\leq 7.$$
The dimension of $\mathcal{C}_{(j_1,j_2,j_3,j_4)}$ is equal to $k=7\cdot2^l-2^{l-2}(j_1+j_2+j_3+j_4)$.
By Theorems 3.3 and 3.4, $\mathcal{C}_{(j_1,j_2,j_3,j_4)}$ is
monomially equivalent to the matrix-product code $[C_6,C_5,C_4,C_3,C_2,C_1,C_0]\cdot A$,
where $C_s=\langle g_s(x)\rangle$, $0\leq s\leq 6$, which is a
negacyclic code over $\mathbb{F}_7$ of length $2^l$
with generator polynomial $g_s(x)=\prod_{j_t>s,  1\leq t\leq 4}f_{j_t}(x)$, and
$A=\left[\begin{array}{ccccccc} 1 & 1& 1& 1& 1& 1& 1  \cr
                                  6 & 5& 4& 3& 2& 1& 0  \cr
                                  1 & 3& 6& 3& 1& 0& 0  \cr
                                  6 & 3& 4& 1& 0& 0& 0\cr
                                  1 & 5& 1& 0& 0& 0& 0\cr
                                  6 & 1& 0& 0& 0& 0& 0\cr
                                  1 & 0& 1& 0& 0& 0& 0  \end{array}\right]$ is a NSC matrix over $\mathbb{F}_7$.
By Theorem 2.1, the minimum Hamming distance of $\mathcal{C}_{(j_1,j_2,j_3,j_4)}$ is equal to
$d={\rm mid}\{7d_6,6d_5,5d_4,4d_3,3d_2,2d_1,d_0\mid d_s\neq 0, \ 0\leq s\leq 6\}$,
where $d_s$ is the minimum Hamming distance of $C_s$ satisfying
$$d_s=\left\{\begin{array}{ll} 1, & {\rm if} \ g_s(x)=1; \cr
                             3, & {\rm if} \ g_s(x)=f_i(x), \ 1\leq i\leq 4; \cr
                             3,   & {\rm if} \
                             g_s(x)=f_i(x)f_j(x), \ (i,j)=(1,2), (3,4); \cr
                             5,   & {\rm if} \
                             g_s(x)=f_i(x)f_j(x), \ (i,j)=(1,3), (1,4), (2,3), (2,4); \cr
                             7, & {\rm if} \ g_s(x)=f_i(x)f_j(x)f_k(x), \ 1\leq i< j<k\leq 4; \cr
                             0,& {\rm if} \ g_s(x)=f_1(x)f_2(x)f_3(x)f_4(x). \end{array}\right.$$
for $s=0,1,\ldots,6$. From
this and by $C_6\supseteq C_5\supseteq C_4\supseteq C_3\supseteq C_2\supseteq C_1\supseteq C_0$, we deduce
that
$d\in\{1,2,3,4,5,6,7,9,10,12,14,15,18,20,21,25,28,30,35$, $42,49,0\}.$
   Now, let $l=3$. In the following table, $N_d$ is the number of negacyclic
codes over $\mathbb{F}_7$ of length $56$ with minimum Hamming
distance $d$, $k_d$ is the possible value for the dimension of a code with minimum Hamming
distance $d$ and $e_{(d,k_d)}$ is the number of negacyclic
codes over $\mathbb{F}_7$ of length $56$ with minimum Hamming
distance $d$ and dimension $k_d$.
\begin{center}
\begin{tabular}{l|l|l}\hline
$d$   & $N_d$  & $k_d^{e_{(d,k_d)}}$  \\ \hline
$2$   & $15$   & $48^1$, $50^4$, $52^6$, $54^4$ \\
$3$   & $175$ &  $28^2$, $ 30^4$, $ 32^6$, $ 34^8$, $ 36^{10}$, $ 38^{12}$, $ 40^{15}$, $ 42^{20}$, $ 44^{24}$, $ 46^{28}$, $ 48^{26}$, $ 50^{16}$\\
& & $ 52^4$, $ 52^4$\\
$4$   & $161$ & $32^1$, $ 34^4$, $ 36^{10}$, $ 38^{20}$, $ 40^{30}$, $ 42^{36}$, $ 44^{32}$, $ 46^{20}$, $ 48^8$         \\
$5$   & $483$ &  $24^1$, $ 26^4$, $ 28^{14}$, $ 30^{28}$, $ 32^{46}$, $ 34^{64}$, $ 36^{78}$, $ 38^{84}$, $ 40^{76}$, $ 42^{52}$, $ 44^{28}$, $ 46^8$        \\
$6$   & $765$ & $16^1$, $ 18^4$, $ 20^{10}$, $ 22^{20}$, $ 24^{36}$, $ 26^{60}$, $ 28^{84}$, $ 30^{104}$, $ 32^{114}$, $ 34^{112}$, $ 36^{96}$        \\
& & $ 38^{72}$, $40^{40}$, $ 42^{12}$\\
$7$   & $1417$ & $8^1$, $ 10^4$, $ 12^{10}$, $ 14^{24}$, $ 16^{46}$, $ 18^{76}$, $ 20^{110}$, $ 22^{144}$, $ 24^{174}$, $ 26^{192}$, $ 28^{188}$\\
& & $ 30^{164}$, $ 32^{128}$, $ 34^{88}$, $ 36^{52}$, $ 38^{16}$         \\
$9$   & $78$ &  $20^2$, $ 22^8$, $ 24^{12}$, $ 26^{12}$, $ 28^{12}$, $ 30^{12}$, $ 32^{12}$, $ 34^8$        \\
$10$   & $36$ &   $24^4$, $ 26^8$, $ 28^8$, $ 30^8$, $ 32^8$       \\
$12$   & $122$ &   $16^2$, $ 18^8$, $ 20^{16}$, $ 22^{20}$, $ 24^{20}$, $ 26^{20}$, $ 28^{20}$, $ 30^{16}$       \\
$14$   & $260$ &   $12^ 4$, $ 14^ {12}$, $ 16^ {24}$, $ 18^ {36}$, $ 20^ {44}$, $ 22^ {48}$, $24^ {44}$, $ 26^{32}$, $ 28^{16}$       \\
$15$   & $130$ &   $12^2$, $ 14^8$, $ 16^ {16}$, $ 18^ {20}$, $ 20^ {24}$, $ 22^{28}$, $ 24^ {24}$, $ 26^ 8$       \\
$18$   & $146$ &  $8^ 2$, $ 10^ 8$, $ 12^ {16}$, $ 14^ {24}$, $ 16^ {28}$, $ 18^ {28}$, $ 20^ {24}$, $ 22^ {16}$        \\
$20$   & $36$ &      $16^ 4$, $ 18^ {16}$, $ 20^ {16}$    \\
$21$   & $194$ &   $4^ 2$, $ 6^ 8$, $ 8^ {16}$, $ 10^ {28}$, $ 12^ {40}$, $ 14^ {44}$, $ 16^ {40}$, $ 18^ {16}$       \\
$25$   & $12$ &    $12^ 4$, $ 14^ 8$      \\
$28$   & $20$ &      $8^ 4$, $ 10^ 8$, $ 12^ 8$    \\
$30$   & $12$ &        $8^ 4$, $ 10^ 8$  \\
$35$   & $24$ & $4^4$, $6^{12}$, $8^8$  \\
$42$   & $4$ &  $4^4$        \\
$49$   & $4$ &  $2^4$        \\
\hline
\end{tabular}
\end{center}

\par
  Precisely, the minimum Hamming distances of all $4094$ nontrivial negacyclic
codes of length $7\cdot2^l$ over $\mathbb{F}_7$ for any integer $l\geq 3$ are given by
Appendix.

\vskip 3mm \noindent {\bf Acknowledgments}
 Part of this work was done when Yonglin Cao was visiting Chern Institute of Mathematics, Nankai University, Tianjin, China. Yonglin Cao would like to thank the institution for the kind hospitality. This research is
supported in part by the National Natural Science Foundation of
China (Grant Nos. 11671235, 11471255, 61571243).

\vskip 3mm \noindent {\bf Appendix. Nontrivial negacyclic
codes of length $7\cdot2^l$ over $\mathbb{F}_7$}

\vskip 3mm \noindent
In the following, $\mathcal{C}_{(j_1,j_2,j_3,j_4)}=\langle f_1(x)^{j_1}f_2(x)^{j_2}f_3(x)^{j_3}f_4(x)^{j_4}\rangle$ is
a negacyclic
codes of length $7\cdot2^l$ over $\mathbb{F}_7$ and $d$ is its minimal
Hamming distance.

\par
  $\bullet$ $d=2$, if $d_0=0$ and $d_1=1$. There are $15$ codes: $\mathcal{C}_{(j_1,j_2,j_3,j_4)}$, where $j_1,j_2,j_3,j_4\in\{0,1\}$ and $(j_1,j_2,j_3,j_4)\neq(0,0,0,0)$.

\par
  $\bullet$ $d=3$, if $d_0=3$ and $d_1\neq 1$; or $d_0,d_1\neq 1$ and  $d_2=1$. There are $175$ codes: $\mathcal{C}_{(j_1,j_2,j_3,j_4)}$, where $(j_1,j_2,j_3,j_4)$ satisfies one of the following five conditions:

\par
  (i) $(j_1,j_2,j_3,j_4)=(a,0,0,0), (0,a,0,0), (0,0,a,0), (0,0,0,a)$, $2\leq a\leq 7$.

\par
  (ii) $(j_1,j_2,j_3,j_4)=(a,b,0,0), (0,0,a,b)$, where $(a,b)\in \{1,2,3,4,5,6$, $7\}^2\setminus\{(1,1)\}$.

\par
  (iii) $(j_1,j_2,j_3,j_4)=(a,0,b,0), (a,0,0,b), (0,a,b,0), (0,a,0,b)$, where $(a$, $b)=(1,2),(2,1),(2,2)$.

\par
  (iv) $(j_1,j_2,j_3,j_4)=(a,b,c,0), (a,b,0,c), (a,0,b,c), (0,a,b,c)$,
$(a,b,c)\in\{1,2\}^3\setminus\{(1,1,1)\}$.

\par
  (v) $(j_1,j_2,j_3,j_4)\in\{1,2\}^4\setminus\{(1,1,1,1)\}$.

\par
  $\bullet$ $d=4$, if $d_0=0,5,7$, $d_1, d_2\neq 1$ and $d_3=1$. There are $161$ codes: $\mathcal{C}_{(j_1,j_2,j_3,j_4)}$, where  $(j_1,j_2,j_3,j_4)$ satisfies one of the following three conditions:

\par
  (i) $(j_1,j_2,j_2,j_4)=(a,0,b,0), (a,0,0,b), (0,a,b,0), (0,a,0,b)$,
where $(a,b)$ $\in \{1,2,3\}^2\setminus \{1,2\}^2$.

\par
  (ii) $(j_1,j_2,j_2,j_4)=(a,b,c,0), (a,b,0,c), (a,0,b,c), (0,a,b,c)$,
where $(a,b$, $c)\in \{1,2,3\}^3\setminus \{1,2\}^3$.

\par
  (iii) $(j_1,j_2,j_2,j_4)\in\{1,2,3\}^4\setminus \{1,2\}^4$.

\par
  $\bullet$ $d=5$, if $d_0=5$, and $d_1, d_2,d_3\neq 1$; or $d_0=0,7$, $d_1, d_2,d_3\neq 1$ and $d_4=1$. There are $483$ codes: $\mathcal{C}_{(j_1,j_2,j_3,j_4)}$, where  $(j_1,j_2,j_3,j_4)$ satisfies one of the following three conditions:

\par
  (i) $(j_1,j_2,j_2,j_4)=(a,0,b,0), (a,0,0,b), (0,a,b,0), (0,a,0,b)$,
where $(a,b)$ $\in \{1,2,3,4,5,6,7\}^2\setminus \{1,2,3\}^2$.

\par
  (ii) $(j_1,j_2,j_2,j_4)=(a,b,c,0), (a,b,0,c), (a,0,b,c), (0,a,b,c)$,
where $(a,b$, $c)\in \{1,2,3,4\}^3\setminus \{1,2,3\}^3$.

\par
  (iii) $(j_1,j_2,j_2,j_4)\in\{1,2,3,4\}^4\setminus \{1,2,3\}^4$.

\par
  $\bullet$ $d=6$, if $d_0=0,7$, $d_1=3$ and $d_2, d_3,d_4\neq 1$; or $d_0=0,7$, $d_1\neq 1,3$,
$d_2,d_3,d_4\neq 1$ and $d_5=1$. There are $765$ codes: $\mathcal{C}_{(j_1,j_2,j_3,j_4)}$, where $(j_1,j_2,j_3,j_4)$ satisfies one of the following seven conditions:

\par
  (i) $(j_1,j_2,j_3,j_4)=(a,1,1,1), (1,a,1,1), (1,1,a,1), (1,1,1,a)$, where $a=5,6,7$.

\par
  (ii) $(j_1,j_2,j_3,j_4)=(a,b,1,1),(1,1,a,b)$, where $(a,b)\in \{2,3,4,5,6,7\}^2\setminus\{2,3,4\}^2$.

\par
  (iii) $(j_{t_1},j_{t_2},j_{t_3})\in \{(a,1,1),(1,a,1),(1,1,a)\mid a=5,6,7\}$, $j_s=0$ for $s\in \{1,2,3,4\}\setminus\{t_1,t_2,t_3\}$, where $1\leq t_1<t_2<t_3\leq 4$.

\par
  (iv) $(j_1,j_2,j_3,j_4)=(a,b,1,0),(a,b,0,1),(1,0,a,b),(0,1,a,b)$, where $(a,b)$ $\in \{2,3,4,5,6,7\}^2\setminus\{2,3,4\}^2$

\par
  (v) $(j_{t_1},j_{t_2},j_{t_3})=(a,s,b,t),(a,s,t,b),(s,a,b,t),(s,a,t,b)$, where $(a,b)\in \{2,3,4,5\}^2\setminus\{2,3,4\}^2$ and $(s,t)\in \{(1,0),(0,1),(1,1)\}$.

\par
  (vi) $(j_{t_1},j_{t_2},j_{t_3})\in \{2,3,4,5\}^3\setminus\{2,3,4\}^3$, $j_s\in\{0,1\}$ for $s\in \{1,2,3,4\}\setminus\{t_1,t_2,t_3\}$, where $1\leq t_1<t_2<t_3\leq 4$.

\par
  (vii) $(j_1,j_2,j_3,j_4)\in \{2,3,4,5\}^4\setminus\{2,3,4\}^4$.

\par
  $\bullet$ $d=7$, if $d_0=0$, $d_1\neq 1,3$, $d_2,d_3,d_4,d_5\neq 1$ and $d_6=1$; or $d_0=7$, $d_1\neq 1,3$ and $d_2,d_3,d_4,d_5\neq 1$.  There are $1417$ codes: $\mathcal{C}_{(j_1,j_2,j_3,j_4)}$, where $(j_1,j_2,j_3,j_4)$ satisfies one of the following five conditions:

\par
  (i) $(j_1,j_2,j_3,j_4)=(a,1,b,1),(a,1,1,b), (1,a,b,1), (1,a,1,b)$,
where $(a,b)$ $\in \{2,3,4,5,6\}^2\setminus \{2,3,4,5\}^2$.

\par
  (ii) $(j_1,j_2,j_3,j_4)=(a,b,c,1),(a,b,1,c), (a,1,b,c), (1,a,b,c)$,
where $(a,b$, $c)\in \{2,3,4,5,6\}^3\setminus \{2,3,4,5\}^3$.

\par
  (iii) $(j_1,j_2,j_3,j_4)\in \{2,3,4,5,6\}^4\setminus \{2,3,4,5\}^4$.

\par
  (iv) $(j_1,j_2,j_3,j_4)=(a,s,b,t),(a,s,t,b), (s,a,b,t), (s,a,t,b)$,
where $(a,b)$ $\in \{2,3,4,5,6,7\}^2\setminus \{2,3,4,5\}^2$ and $(s,t)=(1,0),(0,1)$.

\par
  (v) $(j_1,j_2,j_3,j_4)=(a,b,c,0),(a,b,0,c), (a,0,b,c), (0,a,b,c)$,
where $(a,b$, $c)\in \{2,3,4,5,6,7\}^3\setminus \{2,3,4,5\}^3$.

\par
  $\bullet$ $d=9$, if $d_0=0$, $d_1\neq 1,3$, $d_2=3$ and $d_3,d_4,d_5,d_6\neq 1$. There are $78$ codes: $\mathcal{C}_{(j_1,j_2,j_3,j_4)}$, where $(j_1,j_2,j_3,j_4)$ satisfies one of the following four conditions:

\par
  (i) $(j_1,j_2,j_3,j_4)=(2,2,2,7),(2,2,7,2),(2,7,2,2),(7,2,2,2)$.

\par
  (ii) $(j_1,j_2,j_3,j_4)=(a,b,s,t), (s,t,a,b)$,
where $(a,b)\in \{(7,1),(1,7)\}$ and $(s,t)\in \{(2,1),(1,2)\}$.

\par
  (iii) $(j_1,j_2,j_3,j_4)=(a,b,2,2), (a,2,b,2), (a,2,2,b), (2,a,b,2), (2,a,2,b)$, $(2,2,a,b)$,
where $(a,b)\in \{(7,1),(1,7)\}$.

\par
  (iv) $(j_1,j_2,j_3,j_4)=(a,b,s,t),(s,t,a,b)$, where
$(a,b)\in \{3,4,5,6,7\}^2\setminus\{3,4,5,6\}^2$ and  $(s,t)\in\{(1,2),(2,1),(2,2)\}$.

\par
  $\bullet$ $d=10$, if $d_0=0$, $d_1=5$, $d_2\neq 1,3$ and $d_3,d_4,d_5,d_6\neq 1$.
There are $36$ codes: $\mathcal{C}_{(j_1,j_2,j_3,j_4)}$,
  $(j_1,j_2,j_3,j_4)=(a,1,b,1), (a,1,1,b), (1,a,b,1), (1,a,1,b)$,
where $(a,b)\in \{3,4,5,6,7\}^2\setminus\{3,4,5,6\}^2$.

\par
  $\bullet$ $d=12$, if $d_0=0$, $d_1,d_2\neq 1,3$, $d_3=3$ and $d_4,d_5,d_6\neq 1$.
There are $122$ codes: $\mathcal{C}_{(j_1,j_2,j_3,j_4)}$, where $(j_1,j_2,j_3,j_4)$ satisfies one of the following four conditions:

\par
  (i) $(j_1,j_2,j_3,j_4)=(7,3,3,3), (3,7,3,3), (3,3,7,3), (3,3,3,7)$.

\par
  (ii) $(j_1,j_2,j_3,j_4)=(7,a,b,c), (a,7,b,c), (a,b,7,c), (a,b,c,7)$,
where $(a,b$, $c)\in\{(3,3,s),(3,s,3),(3,3,s)\mid s=1,2\}$.

\par
  (iii) $(j_1,j_2,j_3,j_4)=(a,s,b,t), (a,s,t,b), (s,a,b,t), (s,a,t,b)$,
where $(a,b)$ $\in\{(7,3),(3,7)\}$ and $(s,t)\in\{(1,2),(2,1),(2,2)\}$.

\par
  (iv) $(j_1,j_2,j_3,j_4)=(a,b,s,t), (s,t,a,b)$,
where $(a,b)\in\{4,5,6,7\}^2\setminus\{4,5$, $6\}^2$ and $(s,t)\in\{(1,3),(2,3),(3,3),(3,1),(3,2)\}$.

\par
  $\bullet$ $d=14$, if $d_0=0$, $d_1=7$, $d_2,d_3\neq 1,3$ and $d_4,d_5,d_6\neq 1$.
There are $260$ codes: $\mathcal{C}_{(j_1,j_2,j_3,j_4)}$,
where $(j_1,j_2,j_3,j_4)$ satisfies one of the following two conditions:

\par
  (i) $(j_1,j_2,j_3,j_4)=(a,b,c,1), (a,b,1,c), (a,1,b,c), (1,a,b,c)$,
where $(a,b,c)$ $\in \{4,5,6,7\}^3\setminus\{4,5,6\}^3$.

\par
  (ii) $(j_1,j_2,j_3,j_4)=(a,s,b,t), (a,s,t,b), (s,a,b,t), (s,a,t,b)$,
where $(a,b)\in \{4,5,6,7\}^2\setminus\{4,5,6\}^2$ and $(s,t)\in\{(1,2),(1,3),(2,1),(3,1)\}$.

\par
  $\bullet$ $d=15$, if $d_0=d_1=0$, $d_2,d_3\neq 1,3$, $d_4=3$ and $d_5,d_6\neq 1$.
There are $130$ codes: $\mathcal{C}_{(j_1,j_2,j_3,j_4)}$, where $(j_1,j_2,j_3,j_4)$ satisfies one of the following five conditions:

\par
   (i) $(j_1,j_2,j_3,j_4)=(a,2,b,2), (a,2,b,2), (2,a,b,2), (2,a,2,b)$,
where $(a,b)$ $\in\{(7,4),(7,5),(7,6),(7,7)$, $(4,7),(5,7),(6,7)\}$.

\par
   (ii) $(j_1,j_2,j_3,j_4)=(7,4,4,4), (4,7,4,4), (4,4,7,4), (4,4,4,7)$.

\par
  (iii) $(j_1,j_2,j_3,j_4)=(a,b,4,4), (a,4,b,4), (a,4,4,b), (4,a,b,4),(4,a,4,b)$, $(4,4,a,b)$,
where $(a,b)\in\{(7,2),(7,3),(2,7),(3,7)\}$.

\par
  (iv) $(j_1,j_2,j_3,j_4)=(a,b,s,t), (s,t,a,b)$,
where $(a,b)\in\{(7,5),(7,6),(7$, $7), (5,7),(6,7)\}$ and $(s,t)\in\{(4,2),(4,3),(4,4),(2,4),(3,4)\}$.

\par
  (v) $(j_1,j_2,j_3,j_4)=(a,s,b,t), (a,s,t,b), (s,a,b,t), (s,a,t,b)$,
where $(a,b)\in\{(7,4),(4,7)\}$ and $(s,t)\in\{(3,3),(3,3),(2,3)\}$.

\par
  $\bullet$ $d=18$, if $d_0=d_1=0$, $d_2,d_3,d_4\neq 1,3$, $d_5=3$ and $d_6\neq 1$.
There are $146$ codes: $\mathcal{C}_{(j_1,j_2,j_3,j_4)}$, where $(j_1,j_2,j_3,j_4)$ satisfies one of the following four conditions:

\par
   (i) $(j_1,j_2,j_3,j_4)=(7,5,5,5), (5,7,5,5), (5,5,7,5), (5,5,5,7)$.

\par
  (ii) $(j_1,j_2,j_3,j_4)=(7,a,b,c), (a,7,b,c), (a,b,7,c), (7,a,b,c)$,
where $(a,b$, $c)\in\{(5,5,t),(5,t,5),(t,5,5)\mid t=2,3,4\}$.

\par
  (iii) $(j_1,j_2,j_3,j_4)=(a,s,b,t), (a,s,t,b), (s,a,b,t), (s,a,t,b)$,
where $(a,b)$ $\in\{(7,5),(5,7)\}$ and $(s,t)\in\{2,3,4\}^2\setminus\{(2,2)\}$.

\par
  (iv) $(j_1,j_2,j_3,j_4)=(a,b,s,t), (s,t,a,b)$,
where $(a,b)\in\{(7,6),(7,7),(6$, $7)\}$ and $(s,t)\in\{(5,2),(5,3),(5,4),(5,5),(2,5),(3,5),(4,5)\}$.

\par
  $\bullet$ $d=20$, if $d_0=d_1=0$, $d_2=0,7$, $d_3=5$, $d_4,d_5\neq 1,3$ and $d_6\neq 1$.
There are $36$ codes: $\mathcal{C}_{(j_1,j_2,j_3,j_4)}$, where
$(j_1,j_2,j_3,j_4)=(a,s,b,t),(a,s,t,b),(s,a,b,t)$, $(s,a,t,b)$,
$(a,b)\in\{(7,6),(7,7),(6,7)\}$ and $(s,t)\in\{(2,3),(3,3),(3,2)\}$.

\par
  $\bullet$ $d=21$, if $d_0=d_1=0$, $d_2=7$, $d_3\neq 1,3,5$, $d_4,d_5\neq 1,3$ and $d_6\neq 1$;
or $d_0=d_1=d_2=0$, $d_3\neq 1,3,5$, $d_4,d_5\neq 1,3$ and $d_6=3$.
There are $194$ codes: $\mathcal{C}_{(j_1,j_2,j_3,j_4)}$, where $(j_1,j_2,j_3,j_4)$ satisfies one of the following six conditions:

\par
  (i) $(j_1,j_2,j_3,j_4)=(a,b,c,2), (a,b,2,c), (a,2,b,c), (2,a,b,c)$,
where $(a,b,$ $c)\in \{6,7\}^3\setminus\{(6,6,6)\}$.

\par
   (ii) $(j_1,j_2,j_3,j_4)=(a,s,b,t), (a,s,t,b), (s,a,b,t), (s,a,t,b)$,
where $(a,b)\in \{(7,6),(7,7),(6,7)\}$ and $(s,t)\in\{(2,4),(2,5),(4,2),(5,2)\}$.

\par
   (iii) $(j_1,j_2,j_3,j_4)=(7,6,6,6), (6,7,6,6), (6,6,7,6), (6,6,6,7)$.

\par
  (iv) $(j_1,j_2,j_3,j_4)=(7,a,b,c), (a,7,b,c), (a,b,7,c), (a,b,c,7)$,
where $(a,b$, $c)\in \{(6,6,t),(6,t,6),(t,6,6)\mid t=3,4,5\}$.

\par
  (v) $(j_1,j_2,j_3,j_4)=(a,s,b,t), (a,s,t,b), (s,a,b,t), (s,a,t,b)$£¬
where $(a,b)\in \{(7,6),(6,7)\}$ and $(s,t)\in\{3,4,5\}^2\setminus\{(3,3)\}$.

\par
  (vi) $(j_{t_1},j_{t_2},j_{t_3})=(7,7,s,t), (s,t,7,7)$,
where $(s,t)\in\{(6,3),(6,4),(6,5)$, $(6,6),(3,6),(4,6),(5,6)\}$.

\par
  $\bullet$ $d=25$, if $d_0=d_1=d_2=0$, $d_3\neq 1,3,5$, $d_4=5$ and $d_5,d_6\neq 1,3$.
There are $12$ codes: $\mathcal{C}_{(j_1,j_2,j_3,j_4)}$, where
$(j_1,j_2,j_3,j_4)=(7,s,7,t),(7,s,t,7),(s,7,7$, $t), (s,7,t,7)$,
$(s,t)\in\{(4,3),(3,4),(4,4)\}$.

\par
  $\bullet$ $d=28$, if $d_0=d_1=d_2=0$, $d_3=7$, $d_4\neq 1,3,5$ and $d_5,d_6\neq 1,3$.
There are $20$ codes: $\mathcal{C}_{(j_1,j_2,j_3,j_4)}$, where $(j_1,j_2,j_3,j_4)$ satisfies one of the following two conditions:

\par
  (i) $(j_1,j_2,j_3,j_4)=(7,7,7,3), (7,7,3,7), (7,3,7,7), (3,7,7,7)$.

\par
  (ii) $(j_1,j_2,j_3,j_4)=(7,s,7,t),(7,s,t,7),(s,7,7,t), (s,7,t,7)$,
where $(s,t)$ $\in\{(3,5),(3,6),(5,3),(6,3)\}$.

\par
  $\bullet$ $d=30$, if $d_0=d_1=d_2=d_3=0$, $d_4\neq 1,3,5$, $d_5=5$ and $d_6\neq 1,3$.
There are $12$ codes: $\mathcal{C}_{(j_1,j_2,j_3,j_4)}$, where
$(j_1,j_2,j_3,j_4)=(7,s,7,t)$, $(7,s,t,7),(s,7,7, t), (s,7,t,7)$,
$(s,t)\in\{(4,5),(5,5),(5,4)\}$.

\par
  $\bullet$ $d=35$, if $d_0=d_1=d_2=d_3=0$, $d_4=7$, $d_5\neq 1,3,5$ and $d_6\neq 1,3$;
or $d_0=d_1=d_2=d_3=d_4=0$, $d_5\neq 1,3,5$ and $d_6=5$.
There are $24$ codes: $\mathcal{C}_{(j_1,j_2,j_3,j_4)}$, where $(j_1,j_2,j_3,j_4)$ satisfies one of the following two conditions:

\par
  (i) $(j_1,j_2,j_3,j_4)=(7,7,7,4), (7,7,4,7), (7,4,7,7), (4,7,7,7)$.

\par
  (ii) $(j_1,j_2,j_3,j_4)=(7,s,7,t),(7,s,t,7),(s,7,7,t), (s,7,t,7)$,
where $(s,t)$ $\in\{(4,6),(5,6),(6,6),(6,5),(6,4)\}$.

\par
  $\bullet$ $d=42$, if  $(j_1,j_2,j_3,j_4)=(7,7,7,5), (7,7,5,7), (7,5,7,7), (5,7,7,7)$.

\par
  $\bullet$ $d=49$, if  $(j_1,j_2,j_3,j_4)=(7,7,7,6), (7,7,6,7), (7,6,7,7), (6,7,7,7)$.





\end{document}